\documentstyle[sprocl]{article}

\input{psfig}

\def\be{\begin{equation}}
\def\ee{\end{equation}}
\def\bea{\begin{eqnarray}}
\def\eea{\end{eqnarray}}

\begin{document}

\title{PRE-CLUSTER DYNAMICS IN MULTIFRAGMENTATION}

\author{ W. N\"ORENBERG and G. PAPP}

\address{Gesellschaft f\"ur Schwerionenforschung,
D-64220, Darmstadt, Germany
}




\maketitle\abstracts{
The initial production and dynamical expansion of hot spherical nuclei are
examined as the first stage both in the projectile-multifragmentation and in
central collision processes. The initial temperatures, which are necessary 
for entering the adiabatic spinodal region, as well as the minimum 
temperatures and densities, which are reached in the expansion,
significantly differ for hard and soft equations of
state. Additional initial compression, occurring in central collisions
leads most likely to a qualitatively different multifragmentation
mechanism. Recent experimental data are discussed in relation to the 
results of the proposed model.}

\section{Introduction}

Spinodal instabilities as a possible mechanism of multifragmentation was first
suggested by Bertsch, Siemens and Cugnon~\cite{BeSi83,Cu84}. Recent data from
the ALADIN and FOPI experiments~\cite{Her93,Jeo94,Hsi94,Poc95} shed some light
on such a possible explanation. In particular the discovery of the caloric
curve~\cite{Poc95}, which looks like the one for a liquid-gas phase
transition in water, suggests to study the mechanism of spinodal 
decomposition in more detail.
  
The expansion of the excited (projectile) residue can be regarded as the first
stage in the multifragmentation process. For large enough initial
excitation energies and temperatures the nuclear matter of the residue
reaches states
of mechanical volume instability, where small density fluctuations grow
exponentially \cite{kn:bur92,kn:bat93,kn:pg}. For an illustrative analysis
of the
instability evolution we refer to ref. \cite{kn:col94}. The fragmentation
process itself has been studied within molecular dynamics
\cite{kn:pand85,kn:pand867,kn:paul91,kn:dor94}.
Studies by Friedman \cite{kn:fri1,kn:fri2}
within an expanding emitting source model (EES) show that intermediate
mass fragments (IMFs) are indeed created within time intervals of about
50 fm/c indicating a simultaneous breakup of the residue.

We have been investigating, already for some time now~\cite{PaNo91}, the
expansion dynamics of spherical hot nuclei for peripheral collisions. For 
low initial excitation energies
one obtains monopole vibration with the largest possible amplitudes.
Around an initial temperature of 5 MeV (i.e. $E^*/A = 2$ MeV) the
matter inside the expanding
drop reaches, e.g. in the $T,\varrho$ plane, the boundary of the spinodal
regime. The important feature at higher initial excitation energies is the
occurrence of turning points deep in the spinodal region. The system stays long
enough around these turning points, which all are located around $T=4$ MeV, such
that instabilities can develop and fragments can be formed. At initial
temperatures $T >$ 12 MeV the system runs through the freeze-out density of
$0.3 \varrho_0$ before reaching a turning point. 

\hspace*{3cm}\begin{figure}[t]
\psfig{figure=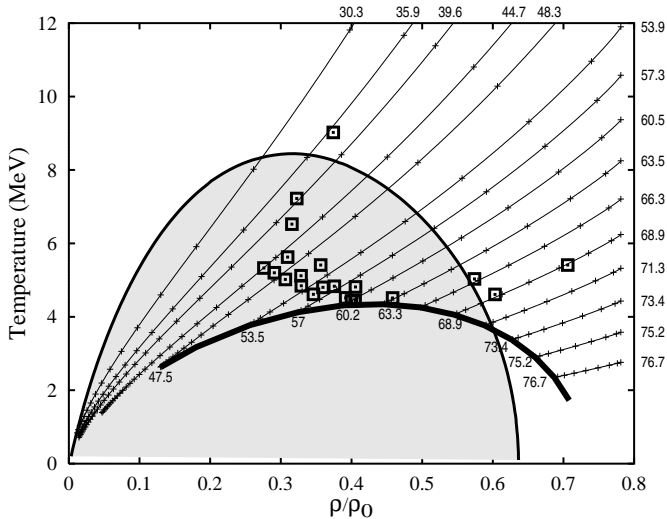,height=6.8cm}
\caption{\label{fig:soft} Time evolution of the Au residue for a soft equation
of state in a peripheral collision (without initial compression) in the 
temperature -- density plane. The crosses on the trajectories indicate 
time steps of 5 fm/c.
The shadowed zone corresponds to the adiabatic mechanical instability region 
for infinite nuclear matter.  The numbers
give the charge numbers initially and at the turning points which are connected
by the heavy solid line. The open boxes denote the experimental
data \protect\cite{Poc95} (without error bars).}
\end{figure}

\section{The Model}

In this section we describe a few essential features of the model.
A more detailed description of the model will be published elsewhere
\cite{kn:pg2}.

First we study the peripheral collision setup.
We start out from a thermalized spherically symmetric projectile-like nucleus
at ground state density, which has been created in the peripheral high-energy
collision according to the abrasion model \cite{kn:oli,kn:abr,kn:abr1}.
The typical temperature for
such a hot nucleus is not too high, such that
the mean free path of nucleons is still comparable or even larger than the size
of the system \cite{kn:he}.
Thus, we can approximately assume homogeneity in density and temperature for
the expanding nucleus.

During the expansion the hot nucleus evaporates particles. We consider only
neutron and proton evaporation. Deuterons may be considered to arise from
coalescence of nucleons. We follow the evolution of the excited remnant and
plot the trajectories in the temperature -- density plane.

If the excitation energy is high enough, the expansion leads into the
instability region of the equation of state. For not too high excitation
energies the expansion of the homogeneous system stops at a certain density.
Around these
turning points the system has enough time ($\ge$ 30 fm/c,
\cite{kn:bur92,kn:pg}) to develop inhomogeneities in density, and subsequently
decay into many fragments.
The final mass and charge spectra is not given by this model, however some
quantitative comparison with experiments concerning temperature and 
excitation energy (caloric curve) can be made.

For a central collision setup we considered a somewhat similar scenario. Here
the initial expanding source can achieve much higher initial temperatures, and
also one should take into account compression. The initial state
was estimated on the base of the shock model~\cite{Nix79}. Due to the
compression the evolution of such a system is so fast, that the evaporation can
be neglected (the ``evaporated'' particles cannot leave the excited source),
and except very low excitation energies no turning is obtained. The calculation
is stop at the break-up density of $0.3\varrho_0$ obtained from analyzing the
ALADIN data.

\section{Results}

\hspace*{3cm}\begin{figure}[t]
\psfig{figure=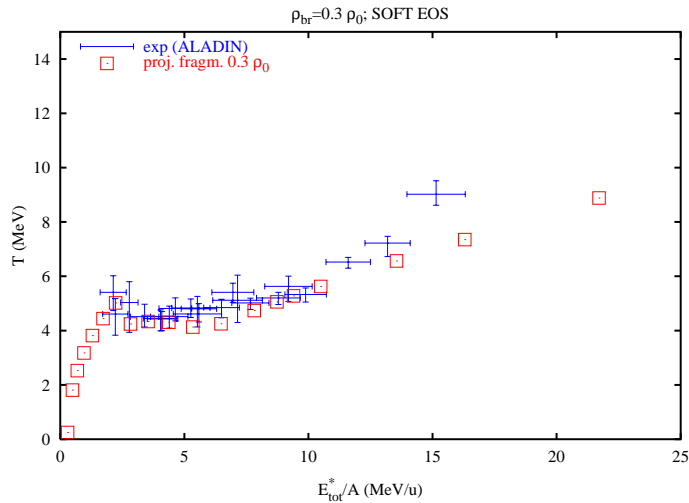,height=6.8cm}
\caption{\label{fig:expal} The caloric curve as deduced by the ALADIN 
collaboration~\protect\cite{Poc95}
(crosses)
compared with the pre-cluster condition.
The total excitation energy
includes flow energy which is significant in the isentropic dynamics for
$E^*/A >$ 15 MeV. The discrepancy at large excitation energies may therefore
indicate that dissipation is not negligible during the expansion 
process. \hfill}
\end{figure}

According to the abrasion model\cite{kn:abr1}, the initial excitation energy
(temperature) is correlated with the charge number of the projectile residue.
These numbers are given on the right and upper parts of Fig.~\ref{fig:soft}.
The calculated expansion trajectories in the ($T,\varrho$)-plane are
significantly different for a soft and a stiff
equation of state (EOS).

We
follow the trajectories up to their turning points if reached within 200 fm/c.
The turning points are indicated by the heavy solid lines. Turning points exist
for initial temperatures up to about 14 MeV. We notice that the turning points
for the same initial excitation lie one to two MeV in temperature and
$0.1 \varrho_0$ to $0.2 \varrho_0$ in density higher for the stiff EOS as
compared to the soft one.
Moreover, we see that the turning points are located almost
at the
same temperatures independent of the initial excitation, i.e. around 4 MeV and
5 MeV for the soft and hard EOS, respectively. This plateau does not move
if one changes the initial excitation energy, the only effect is, that the
charge numbers become different initially and at the turning points.

A natural criterion for multifragmentation to occur, is whether the system
reaches the region of volume instability (adiabatic spinodal region), where the
derivative of the pressure with respect to the density along an adiabat
becomes negative. Since the system is closed we consider the adiabatic process
to be the relevant one. In Fig.~\ref{fig:soft} the region
of instability
is shadowed and bounded by the spinodal. To enter the spinodal region and stay
there for more than 30 fm/c one
needs initial temperatures of about 8 MeV and 11 MeV for the soft and hard
EOS, respectively.

Because of the occurrence of turning points in the expansion,
the multifragmentation process of the projectile residue has a
unique feature:
the subsequent decay into fragments is expected to be rather free
from collective flow. This is different from the fragmentation of compressed
compound systems which are formed in central collisions.
As our calculations show already for a small initial compression (1.5
$\varrho_0$) no turning points occur for realistic
initial excitation energies. Thus the fragmentation process becomes
qualitatively different and looks more like an explosion
\cite{kn:rei}.

In Fig.~\ref{fig:expal} we compare our results for the temperature and the
total excitation energy in the remnant with experimental
data from the ALADIN experiment. Our points are determined at the turning
points if the corresponding density is higher than $\varrho_{br} \approx
0.3 \varrho_0$, or at the break-up density otherwise. For very low excitation
energies the average values of the vibration are chosen.
 There are some deviation at high excitation
energies, which is due to the arising flow energy in our model.

\begin{figure}[t]
\psfig{figure=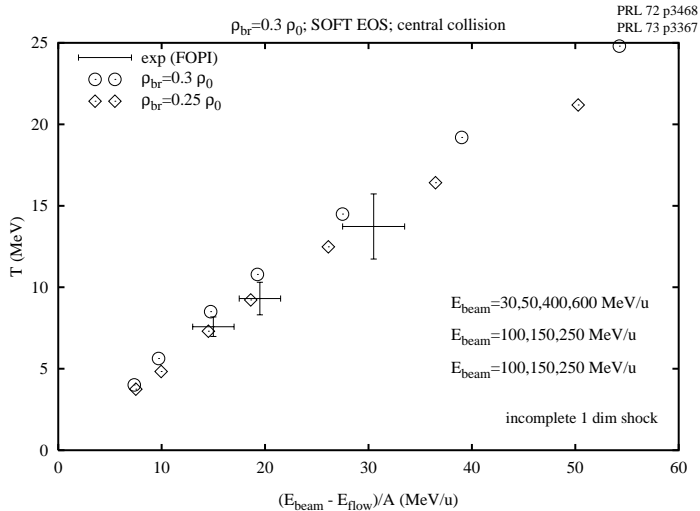,height=6.8cm}
\caption{\label{fig:fopi} The caloric curve as deduced from the FOPI
experiments at 100, 150 and 250 MeV/u collisions\protect\cite{Jeo94,Hsi94}
 (crosses) compared with our
results (circles for $\varrho_{br}$=0.3 $\varrho_0$ and
diamonds for $\varrho_{br}$=0.25 $\varrho_0$). The flow energy is subtracted 
from the excitation energy.}
\end{figure}

In Fig.~\ref{fig:fopi} the comparison with the FOPI data~\cite{Jeo94,Hsi94} are
plotted. The experimental results are consistent with the expansion dynamics 
of an initially compressed nucleus and a break-up density of $0.3\varrho_0$.

Figs.~\ref{fig:expal} and ~\ref{fig:fopi} show another important difference
between peripheral and central collisions. For high excitation energies both
curves follow the $T = \frac{2}{3} E_{kin}/A$ free gas limit shifted by the
remaining binding energy. However at low energies the peripheral setup shows a
phase-transition like behavior: with decreasing excitation energy the
temperature stays constant, and at even lower energies the nuclear liquid
relation is reproduced.

\hspace*{3cm}

We conclude that to a large extent the temperatures observed in
multifragmentation can be understood from the expansion dynamics into the
spinodal regime and across the freeze-out density both for peripheral and for
central collisions, despite the different break-up mechanism. The observed
"phase-transition" plateau depends on the initial compression of the remnant,
and for usual central collision is missing.

\section*{References}

\end{document}